\documentclass{ws-mpla}

\begin{document}

\markboth{Dong-han Yeom}
{Reviews and perspectives on black hole complementarity}

\catchline{}{}{}{}{}

\title{Reviews and perspectives on black hole complementarity\footnote{A proceeding for CosPA2008. Talk on the 29th of October, 2008, Pohang, Korea.}}

\author{\footnotesize Dong-han Yeom}

\address{Department of Physics, KAIST,\\
Daejeon, 305-701, South Korea\\
innocent@muon.kaist.ac.kr}

\maketitle


\begin{abstract}
If black hole complementarity is the correct idea to resolve the information loss problem, it should apply to general black holes. We suggest two models: Frolov, Markov, and Mukhanov's regular black hole and a charged black hole. These models can work as counterexamples to black hole complementarity. It has been mentioned that a large number of massless fields is an important condition to justify these models. The invalidity of this principle may imply that the holographic principle must be re-interpreted; the information loss problem, as well, should be re-considered.

\end{abstract}

\ccode{}

\section{Introduction}

The black hole information loss problem is one of the most important issues in modern physics. If there is information loss, it may imply the fundamental limitation of the quantum theory of gravity. Thus, there is almost a consensus that accepts the conservation of information. The next natural question becomes, \textit{how can information come out of a black hole?}

\paragraph{Motivation and presence of black hole complementarity}

Let us assume two facts: first, the unitarity of quantum mechanics; second, the area of a black hole as its entropy. One may guess that, from the first assumption, a local observer should reconstruct all bits of information. If we accept the second assumption, from the information theoretical consideration, one notices that some information should come out of a black hole even if the black hole is sufficiently large. \textit{Even when a black hole is big, it should emit information by Hawking radiation after the information retention time.} However, we know that free-falling information will touch the singularity, and information should be located at the center of the black hole. Then, the free-falling information is located at the center, and, at the same time, the information is located on the outside via the Hawking radiation. Of course, this is impossible according to the no cloning theorem. Now, black hole complementarity says that this can happen if there is no witness of the violation of natural laws. If the free-falling information and the Hawking radiation cannot communicate forever, there will be essentially no problem. This can resolve the information loss paradox in a fascinating way.

As this author can safely assert, black hole complementarity is a quite general consensus between string theorists. Complementarity was a crucial step toward the speculation that produced the holographic principle. Moreover, during the 1990s, some important techniques were developed to realize the holographic principle: the D-brane picture and AdS/CFT. If AdS/CFT is true, a black hole in anti de Sitter space should be unitary. Also, from the D-brane picture, we understand that the black hole area is proportional to the statistical entropy. Then, because of some arguments from the information theory, black hole complementarity must be implemented in a black hole.

\paragraph{Duplication experiment}

Black hole complementarity is a falsifiable hypothesis. In other words, it assumes that there is no witness of the duplication of information. Is this really true? For a Schwarzschild black hole, this could be checked. First, the outside observer cannot see the duplication, since the Hawking radiation is generated from the event horizon. Second, the inside observer cannot see the duplication, since the inside observer should collapse to the singularity quite quickly.\footnote{To see the duplication, the in-falling information should send a signal to the out-going direction during time $\Delta t \sim \exp(-M^{2})$, where $M$ is the black hole mass. Then, from the uncertainty principle, to send a signal with quantum information during $\Delta t$, it needs energy $\Delta E \sim \exp(M^{2})$. However, it is greater than the black hole's mass $M$ itself, and this thought experiment seems to be impossible.}

However, one can notice that this argument about the duplication experiment assumes some accidental facts: first, it assumes the singularity; second, it assumes that the Hawking radiation is generated at the event horizon. However, these assumptions are not true in general. In this paper, the author will discuss two models that violate one of these assumptions. First, we will drop the first assumption, and discuss a regular black hole. Second, we will drop the second assumption, and discuss a charged black hole.

\section{Constructing counterexamples to black hole complementarity}

\paragraph{Frolov, Markov, and Mukhanov's regular black hole}

\begin{figure}
\begin{center}
\includegraphics[scale=0.5]{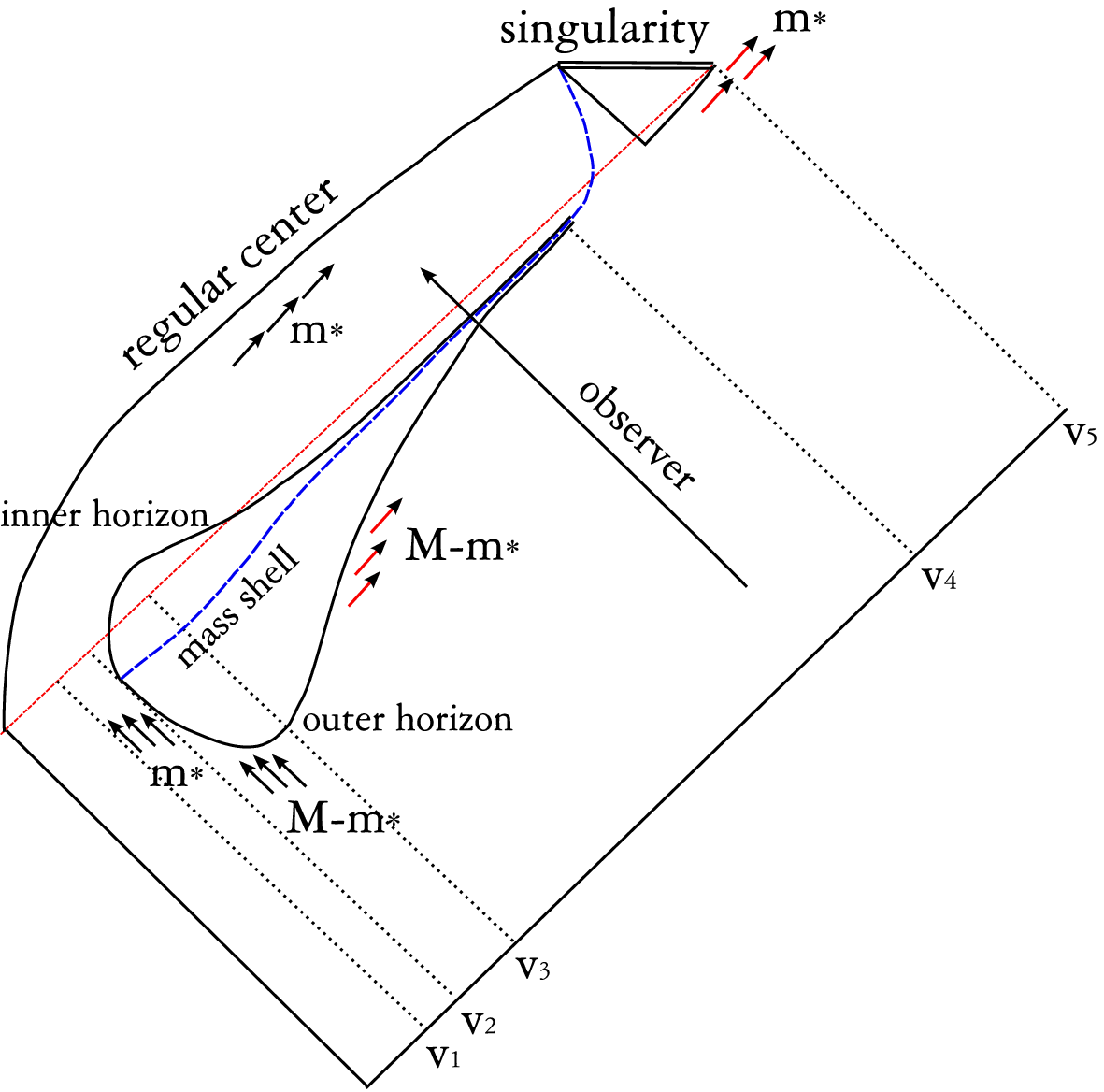}
\includegraphics[scale=0.45]{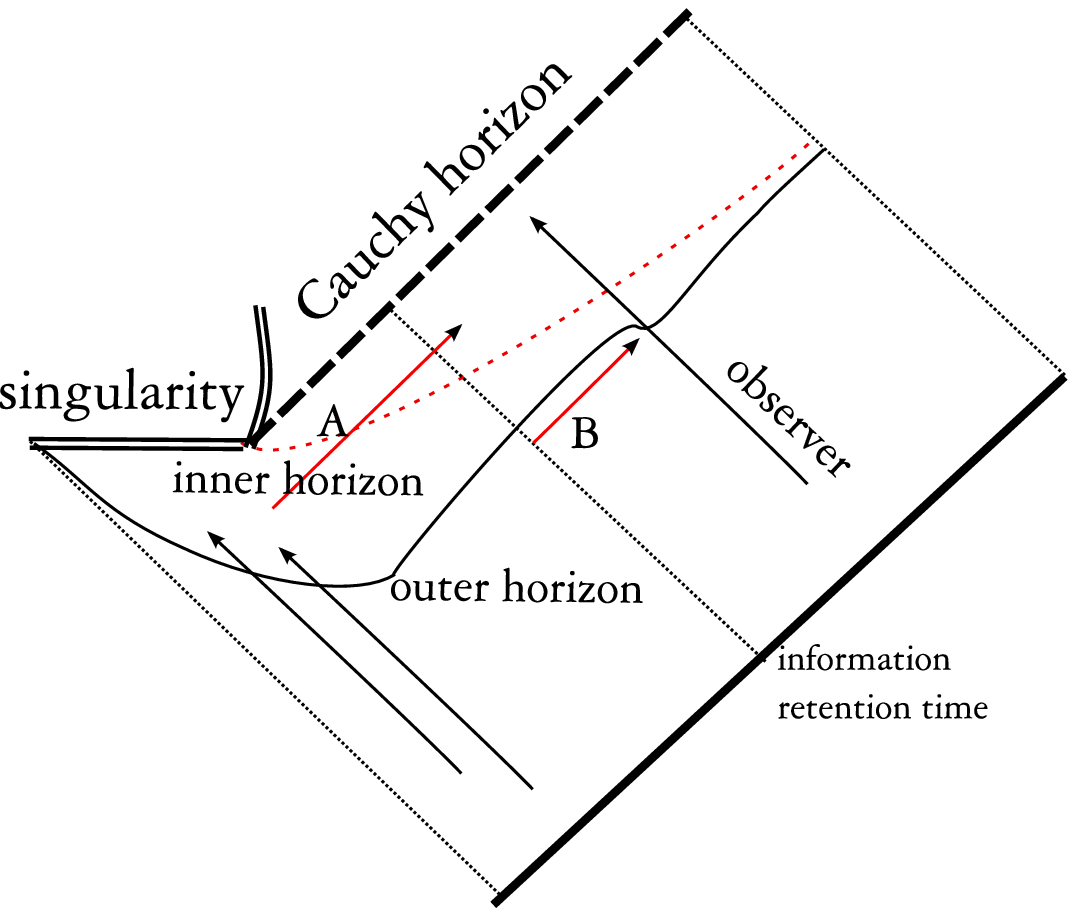}
\caption{\label{fig:penrose} Causal structures of Frolov, Markov, and Mukhanov's regular black hole (left) and a dynamical charged black hole (right).}
\end{center}
\end{figure}

First, let us discuss a regular black hole. Regular means that there is no singularity. From the singularity theorem, it is quite natural to think that there is a singularity inside a black hole. However, if one drops some of assumptions of the theorem, one may allow a regular black hole solution. In general, for static metrics, there is no singularity via the violation of causality conditions. Moreover, even though a static metric holds the null energy condition, for a dynamical metric, it may violate all kinds of energy conditions via the Hawking radiation. Thus, the existence of regular space-time in realistic situations is not so strange.

We will discuss Frolov, Markov, and Mukhanov's model. This model assumes a local false vacuum. Thus, the static metric holds the null energy condition. Since we need to paste the false vacuum and the true vacuum, we need a transition layer by a mass shell. This condition could be calculated. The metric is
\begin{eqnarray} \nonumber
ds^{2} = -(1-2m(r,l)/r) dt^{2}+(1-2m(r,l)/r)^{-1}dr^{2}+r^{2}d\Omega^{2},
\end{eqnarray}
where $m(r,l) = m\theta(r-r_{0})+(r^{3}/2l^{2})\theta(r_{0}-r)$ with mass $m$. Here, $l=(\Lambda/3)^{-1/2}$ is the Hubble scale parameter of the inside de Sitter metric, and $r_{0}=(12/\alpha)^{1/6}(2m/l)^{1/3}l$ is the radius of the false vacuum boundary (we can choose $\alpha$ as a free parameter). Note that the energy condition or the matter configuration is quite realistic, since one needs to assume just a false vacuum. Also, since it uses the thin shell approximation, as long as the approximation is true, its causal structure does not depend on the details of the transition region.

To describe a dynamical case, we can use the Vaidya metric. One can notice that this model has two horizons $r_{+}=2m$ and $r_{-}=l$. In general, this kind of model will approach the extreme limit, and approach a stable remnant with mass $m_{*}=l/2$. However, if the false vacuum collapses to singularity before the extreme limit, it will form a Schwarzschild black hole; this assumption seems to be plausible in realistic situations. Integrating these facts, we can draw the causal structure of the regular black hole model (the left diagram of Figure \ref{fig:penrose}; where $M$ is the maximum mass). The information retention time should be located between $v_{3}$ and $v_{4}$, and the observer who falls into the black hole between the times will compare the Hawking radiation with the free-falling information. Thus a duplication experiment becomes possible.

One potential problem is the inner horizon. Because of the mass inflation, the inner horizon will be unstable in general, and it may form a curvature singularity. However, from some numerical work, we can conclude that we can qualitatively trust the causal structure. Now, I will introduce the scheme of the numerical calculations, as well as a dynamical charged black hole.

\paragraph{Dynamical charged black hole}

To construct a charged black hole model, we assumed a complex massless scalar field $\phi$ that is coupled with the electromagnetic field $A_{\mu}$:
\begin{eqnarray} \nonumber
\mathcal{L} = - (\phi_{;a}+ieA_{a}\phi)g^{ab}(\overline{\phi}_{;b}-ieA_{b}\overline{\phi})-\frac{1}{8\pi}F_{ab}F^{ab},
\end{eqnarray}
where $F_{ab}=A_{b;a}-A_{a;b}$, and $e$ is the unit charge. For convenience, spherical symmetry is a useful assumption: $ds^{2} = -\alpha^{2}(u,v) du dv + r^{2}(u,v) d\Omega^{2}$, where we used the double-null coordinate (our convention is $[u,v,\theta,\phi]$). Since the electromagnetic field is a gauge field, we can fix a gauge $A_{\mu}=(a,0,0,0)$. Then, these assumptions give the Einstein tensor $G_{\mu\nu}$ and the stress-energy tensor $T_{\mu\nu}$ components, as well as equations of motion for the scalar field and the electromagnetic field. To describe the dynamical case, we needed to include the Hawking radiation. For this purpose, we introduced the renormalized stress-energy tensor $\langle T_{\mu\nu} \rangle$ via the S-wave approximation from the 2-dimensional results (which are divided by $4 \pi r^{2}$). This is a reasonable assumption for a spherically symmetric case. Because of spherical symmetry, assigning of initial conditions is not a difficult problem. We pushed a field configuration at the initial surface by choosing a proper function $\phi(u_{i},v)$, where $u_{i}$ represents the initial retarded time. Finally, we solve the Einstein equation $G_{\mu\nu}=8\pi (T_{\mu\nu}+\langle T_{\mu\nu} \rangle)$, as well as field equations for the scalar field and the gauge field.

The right diagram of Figure \ref{fig:penrose} is the causal structure of a dynamical charged black hole. The meaning is clear. Initially, there is a space-like singularity due to the mass-inflation. If there is no Hawking radiation, there will be an inner horizon at $v\rightarrow\infty$, and the inner horizon will be a curvature singularity. However, because of the Hawking radiation, the black hole will become closer to an extreme black hole: the outer horizon bends the time-like direction, and the inner horizon bends the space-like direction. Because of the singularity, there is the Cauchy horizon, which one cannot determine from the initial data. And, there is enough distance between the inner horizon and the Cauchy horizon.

From the calculation, we can notice that the mass function increases exponentially around the inner horizon: $\sim \exp(\kappa_{i}(u+v))$, where $\kappa_{i}$ is the surface gravity of the inner horizon. However, as long as $u$ and $v$ are finite, the mass function or curvature functions are finite. Thus, the inner horizon is regular in the general relativistic sense. Of course, if a curvature function becomes bigger than the Planck scale, it will be a problem. However, in our scheme, \textit{the Planck scale is not determined until we choose a number of massless fields}: when one chooses $c=G=1$, for one specific simulation, $N\hbar$ needs to be fixed as a constant, where $N$ is the number of massless fields. After determining $\hbar$ or $N$, we could determine the Planck scale. If we choose large $N$, then it implies small $\hbar$, and the curvature cutoff becomes bigger and bigger, i.e., via the re-scaling, physical curvatures become smaller and smaller. Therefore, large $N$ will make entire solutions trustable in a semi-classical sense.

Now, we can do a semi-classical thought experiment with our causal structure. One can easily see that there is enough time to send a signal $A$, and the observer can compare both the information of $A$ and $B$.

\paragraph{Assumptions of counterexamples}

For the dynamical charged black hole, we needed to assume two conditions. First, we did not include the pair-creation effects. Thus, we needed to assume that pair-creation effects of electrons are not more dominant than the Hawking radiation. Of course, this can be justified if the mass of the black hole is sufficiently large. And also, if the charge is sufficiently large, the black hole tends to approach the extreme limit. Of course, this assumption is not bad in a physical sense. Second, we needed to regularize the region between the inner horizon and the Cauchy horizon. This region has a large curvature in general. However, we noticed that the large number of massless fields will re-scale all values, and will make the entire region semi-classically convincible.\footnote{One may speculate that if there is an unknown selection principle as a conspiracy between the singularity and the outer horizon, black hole complementarity can be maintained. For example, if one assumes the Horowitz-Maldacena proposal, which is a stronger proposal than complementarity, one may rescue complementarity via quantum teleportation. However, as the author discussed in another place, when one considers the Hayden-Preskill argument, this conspiracy seemed to be improbable: in the Hayden-Preskill case, the out-going information is already selected, and no one can implement further a selection principle by a natural way.} And, as I noticed, the second condition is crucial for the regular black hole case.

Therefore, one of the author's conclusions is that, \textit{if one assumes a large number of massless fields, one allows the violation of black hole complementarity.}\footnote{\textit{Note on a Schwarzschild black hole with large N limit.} Even when we consider a Schwarzschild black hole, if one allows large $N$, black hole complementarity can be violated. When we fix $c=G=1$, by tuning $\hbar = 1/N$, all mass, length, and time scales will be re-scaled by $\sqrt{N}$. Then, $\Delta t \sim \exp(-M^{2})$ should be re-scaled by $\sqrt{N}\exp(-M^{2})$, and from the uncertainty relation, the required energy will be on the order of $\Delta E \sim (1/\sqrt{N})\exp(M^{2})$. As a physical value, $\Delta E > \sqrt{N}M$ is the consistency condition for complementarity. Thus, for sufficiently large $N$, the violation of complementarity can be allowed even in a Schwarzschild black hole.}

\section{Implications and perspectives}

\paragraph{Interpretation: Ontological holography vs. Epistemological holography}

Therefore, if the assumption of a large number of massless fields is physically possible, we can conclude that black hole complementarity is the wrong way to resolve the information loss problem. However, as the author commented, black hole complementarity is a quite natural interpretation whenever we assume the unitarity and the entropy formulas.

If we trust the unitarity of quantum mechanics, one possible (and I think the only) interpretation is that, the area is not the \textit{real} entropy of a black hole. The holographic principle says that physical degrees of freedom of bulk can be mapped to its boundary. However, there is no reason to think that the mapping is physical. It is more reasonable to think that the mapping is just mathematically and apparently true in a certain limit, as the loop quantum gravity argues. Then, we do not need to conclude black hole complementarity, and it resolves an apparent paradox. I will call this new interpretation of holography \textit{the epistemological holography}.

\paragraph{Information loss in black holes}

What about the information loss problem? If the area is not the real information capacity, the Hawking radiation may not contain sufficient information. Then, information will likely be contained by a black hole until the final stage. One possible way is that the final stage contains all of information; the other way is that the information has already spread over space-time, although the observer could not re-construct the initial information. More detailed discussions will be found in the author's future work.

\paragraph{Perspectives}

Finally, I will comment on possible further work.

\begin{itemize}
  \item \textit{Construction of a large number of massless fields.} In a semi-classical sense, large $N$ is not a bad assumption. But, it is still unclear whether string theory allows large $N$. If it is impossible for a fundamental reason, this will give important implications to the information loss problem.
  \item If black hole complementarity is not true, then, as we discussed, the holographic principle should be re-interpreted. Then, \textit{is this new interpretation consistent with known calculations?}
  \item The information loss problem should be re-considered. Is the remnant picture true? If not, why not, and \textit{what is the correct way to resolve the information loss problem?}
  \item It may have implications on de Sitter complementarity and the holographic measure of the multiverse.
\end{itemize}

This talk is based on the following works (and references therein).

\end{document}